\documentclass[11pt]{article}
\usepackage{url}
\usepackage{graphicx}
\usepackage{amsfonts}
\usepackage{amssymb}
\usepackage{latexsym}

\newcommand{\notyet}[1]{}

\newcommand{\ABox}{
\raisebox{3pt}{\framebox[6pt]{\rule{6pt}{0pt}}}
}
\newenvironment{proof}{{\bf Proof:}}{\hfill\ABox}

\newtheorem{theorem}{{\bf Theorem}}

\newtheorem{lemma}[theorem]{Lemma}

\newcommand\R{\mathbb{R}}
\newcommand{\squeezelist}{\setlength{\itemsep}{0pt}}
\newcommand{\hide}[1]{}

\begin{document}

\title{Epsilon-Unfolding Orthogonal Polyhedra}
\author{
Mirela Damian%
   \thanks{Dept. Comput. Sci., Villanova Univ., Villanova,
    PA 19085, USA.
   \protect\url{mirela.damian@villanova.edu}.}
\and
Robin Flatland%
   \thanks{Dept. Comput. Sci., Siena College, Loudonville, NY 12211, USA.
    \protect\url{flatland@siena.edu}.}
\and
Joseph O'Rourke%
    \thanks{Dept. Comput. Sci., Smith College, Northampton, MA
      01063, USA.
      \protect\url{orourke@cs.smith.edu}.
       Supported by NSF Distinguished Teaching Scholars award
       DUE-0123154.}
}
\date{}
\maketitle

\begin{abstract}
An \emph{unfolding} of a polyhedron is produced by cutting 
the surface and flattening 
to a single, connected, planar
piece without overlap 
(except possibly at boundary points).
It is a long unsolved problem to determine whether every
polyhedron may be unfolded.  Here we prove, via an
algorithm, that every \emph{orthogonal polyhedron} (one whose
faces meet at right angles) of genus zero may be unfolded.
Our cuts are not necessarily along edges of the polyhedron,
but they are always parallel to polyhedron edges.
For a polyhedron of $n$ vertices, portions of the
unfolding will be rectangular strips which, in the worst case,
may need to be as thin as
$\varepsilon = 1/2^{\Omega(n)}$.
\end{abstract}


\section{Introduction}
\label{sec:Intro}
Two unfolding problems have remained unsolved for many
years~\cite{do-sfucg-05}: (1)~Can every convex polyhedron be
edge-unfolded? (2)~Can every polyhedron be unfolded?
An \emph{unfolding} of a 3D object is an isometric mapping of its
surface to a single, connected planar piece, the ``net'' for the
object, that avoids overlap. An \emph{edge-unfolding} achieves the
unfolding by cutting edges of a polyhedron, whereas a \emph{general
unfolding} places no restriction on the cuts.
General unfoldings are known for convex polyhedra, but not
for nonconvex polyhedra.
It is known that some nonconvex polyhedra cannot be edge-unfolded,
but no example is known of a
nonconvex polyhedron that cannot be unfolded with unrestricted cuts.
The main result of this paper is that the class of
genus-zero orthogonal polyhedron have a general unfolding.
As we only concern
ourselves with general unfoldings of genus-zero polyhedra in this paper, 
we will drop the
``general'' and ``genus-zero'' modifiers when clear from the context.

The difficulty of the unfolding problem has led to a focus
on \emph{orthogonal polyhedra}---those whose faces meet at angles that
are multiples of $90^\circ$---and especially on genus-zero polyhedra,
i.e., those whose surface is homeomorphic to a sphere.
This line of investigation was initiated
in~\cite{bddloorw-uscop-98}, which established that certain subclasses of
orthogonal polyhedra have an unfolding: 
\emph{orthostacks} and \emph{orthotubes}.
Orthostacks are extruded orthogonal polygons stacked along one
coordinate direction.  The orthostack algorithm does not achieve an
edge unfolding, but it is close, in a sense we now describe.

A \emph{grid unfolding} adds edges to the surface by intersecting
the polyhedron with planes parallel to Cartesian coordinate planes
through every vertex.  This concept has been used to achieve
grid \emph{vertex unfoldings} of orthostacks~\cite{dil-gvuo-04},
and later, grid vertex unfoldings of all genus-zero orthogonal 
polyhedra~\cite{dfo-gvuop-06}.  (A ``vertex unfolding'' is a
loosening of the notion of unfolding that we
do not pause to define~\cite{deeho-vusm-03}.)
A $k_1 \times k_2$ \emph{refinement} of a surface~\cite{do-op03-04}
partitions each face further into a $k_1 \times k_2$ grid of faces;
thus  a $1 \times 1$ refinement is an unrefined
grid unfolding.
The orthostack algorithm achieves a $2 \times 1$ refined grid unfolding.
It remains open to achieve a grid unfolding of orthostacks.
(It is known that not all orthostacks may be edge unfolded.)

The algorithm we present in this paper could be characterized
as achieving a $2^{O(n)} \times 2^{O(n)}$ refined grid unfolding
of orthogonal polyhedra of $n$ vertices.
We coin the term \emph{epsilon-unfolding} to indicate a refinement
with no constant upper bound, but which instead grows with $n$.
In our case, some portions of the unfolding might be
$\varepsilon$-thin, with $\varepsilon = 1/2^{\Omega(n)}$. 

Our algorithm has it roots in the staircase unfolding
of~\cite{bddloorw-uscop-98}, in the spiral strips used
in~\cite{dfo-umt-05}, and the band structure exploited
in~\cite{dfo-gvuop-06}, but introduces several new ideas,
most notably a recursive spiraling
pattern whose nesting leads to the $\varepsilon$-thin characteristic of
the unfolding.

\subsection{Definitions}
Let $O$ be a solid, genus-zero, orthogonal polyhedron.
We assume $O$ has all edges parallel to
$xyz$ axes of a Cartesian coordinate system.
We use the following notation to describe the six types of faces
of $O$,
depending on the direction in which the outward normal points:
{\em front}: $-y$; {\em back}: $+y$; {\em left}: $-x$;
{\em right}: $+x$; {\em bottom}: $-z$; {\em top}: $+z$.
We take the $z$-axis to define the vertical direction.
The spiral paths that play a key role in our algorithm will wrap
around \{top, right, bottom, left\} faces, and ``move'' in the
front/back $y$-direction.

\begin{figure}[htbp]
\centering
\includegraphics[width=0.5\linewidth]{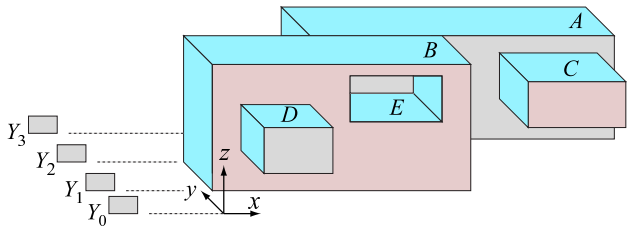}
\caption{Definitions: $A$, $B$, $C$, and $D$ are protrusions; $E$ is
a dent.} \label{fig:defs}
\end{figure}

Let $Y_i$ be the plane
$y=y_i$ orthogonal to the $y$-axis. Let $Y_0, Y_1, \ldots, Y_i,
\ldots$ be a finite sequence of parallel planes passing through
every vertex of $O$, with $y_0 < y_1 < \cdots < y_i < \cdots$. We
call the portion of $O$ between planes $Y_i$ and
$Y_{i+1}$ \emph{layer $i$};
it includes a collection of disjoint connected
components of $O$. 
We call each such component a {\em slab}.
Referring to Figure~\ref{fig:defs}, layer $0$, and $2$ each
contain one slab ($D$ and $A$, respectively), whereas
layer $1$ contains two slabs ($B$ and $C$). The surface piece
that surrounds a slab is called a {\em band}
(labeled in Figure~\ref{fig:defs}). 
Each band has two \emph{rims}, the cycle of edges that lie
in its bounding $Y_i$ and
$Y_{i+1}$ planes.
Each slab is bounded by an outer
band, but it may also contain inner bands, bounding holes.
Outer bands are called \emph{protrusions} and inner bands are called
\emph{dents} ($E$ in Figure~\ref{fig:defs}).

\subsection{Dents vs. Protrusions}
As we observed in~\cite{dfo-gvuop-06}, 
dents may be treated exactly the same as protrusions
with respect to unfolding,
because unfolding of a $2$-manifold to another surface
(in our case, a plane)
depends only on the intrinsic geometry of the surface, and not on how
it is embedded in $\R^3$.
Note that we are only concerned with
the final unfolded ``flat state''~\cite{do-sfucg-05}, and not about
possible intersections during a continuous sequence of partially
unfolded intermediate states.
All that matters for unfolding is which faces share an edge,
and the cyclic ordering of the faces incident to a vertex, i.e.,
our unfolding algorithms will make local decisions and will be oblivious
to the embedding in $\R^3$.
These local relationships are identical if we conceptually ``pop-out''
dents to become protrusions
(this popping-out is
conceptual only, for it could produce self-intersecting objects.)
Henceforth, we will describe only protrusions in our algorithms,
with the understanding that nothing changes for dents.
This shows that our algorithm works on a wider class
of objects than the orthogonal polyhedra,
an observation we do not pursue. 

\subsection{Overview}
The algorithm first partitions the polyhedron $O$ by the $Y_i$ planes,
and then forms an ``unfolding tree'' $T_U$ whose nodes are bands, and with
a parent-child arc representing a ``$z$-beam'' of visibility in their
shared $Y_i$ plane that connects the bands.  
Front and back children 
are distinguished according to the relative $y$-positions of
the children with respect to the parent.
The recursion follows a preorder traversal of
this tree. A thin spiral path winds around the
\{top, right, bottom, left\} faces of a root band $b$, visits each
of the front children recursively, and then each of the back children
recursively.  
The children are visited in a parentheses-nesting order that is forced
by the turn-around requirements.
At all times the spiral alternates turns
so that its unfolding to the plane is a 
staircase-like path monotone with respect to the horizontal.
(Cf. Figures~\ref{fig:box1}, \ref{fig:box2}, \ref{fig:boxfill1}, \ref{fig:complete.example}.)
When the path finishes spiraling around the last back child of $b$,
it is deeply nested inside the spiral, and must retrace the entire
path to return adjacent to its starting point.  (It is this retracing,
recursively encountered, that causes the exponential thinness.)
Again this is accomplished while maintaining the staircase-like layout.
Finally, the front and back faces are hung above and below
the staircase. Following
the physical model of cutting out the net from a sheet of paper,
we permit cuts representing \emph{edge overlap}, where the boundary
touches but no interior points overlap. This can occur 
when hanging the front and back faces
(cf.~Figures~\ref{fig:boxfill1}, \ref{fig:complete.example}).

\section{Unfolding Extrusions}
\label{sec:unfolding.extrusions}

It turns out that nearly all algorithmic issues are present in
unfolding polyhedra that are extrusions of simple orthogonal
polygons. Therefore we will describe the algorithm for this simple
shape class first, in detail, and then show that the ideas extend directly to
unfolding all orthogonal polyhedra.

\begin{figure}[htbp]
\centering
\includegraphics[width=0.65\linewidth]{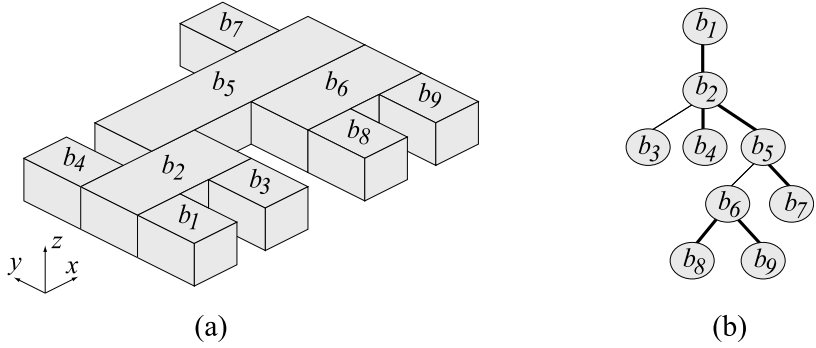}
\caption{(a) Partition of $O$'s $x$ and $z$ perpendicular faces into
bands. (b) Unfolding tree $T_U$. Thin arcs connect a parent to its
front children; thick arcs connect it to its back children. }
\label{fig:extrusion.partition}
\end{figure}

Let $O$ be a polyhedron that is an extrusion in the $z$ direction of
a simple orthogonal polygon. We start with the partition $\pi$ of
$O$ induced by the $Y_i$ planes 
passing through every vertex, as
described in Section~\ref{sec:Intro}. Each element in the partition is 
box surrounded by a four-face band.
The dual graph of $\pi$ has a node for each
band and an edge between each pair of adjacent bands.  For ease of
presentation, we will use the terms {\em node} and {\em band}
interchangeably.  Because $O$ is simply connected, the dual graph is
a tree $T_U$, which we refer to as the {\em unfolding tree}. The
root of $T_U$ is any band that intersects $Y_0$. See
Figure~\ref{fig:extrusion.partition}.
%

We distinguish between the two rims of each band via a recursive
classification scheme. The rim of the root band at $y_0$ is the
\emph{front} rim; the other one is the {\em back} rim.  For any
other band $b$, the rim of $b$ adjacent to its parent is the
{\em front} rim, and the other is the {\em back} rim.  
In Figure~\ref{fig:extrusion.partition}a, for example,
the front rim of $b_8$ is at $y_1$.
A child is a
{\em front child} ({\em back child}) if it is adjacent to the front
(back) rim of its parent.  In Figure~\ref{fig:extrusion.partition}b,
thin arcs connect a parent to its front children; thick arcs connect
a parent to its back children.

In the following we describe the recursive unfolding algorithm.  We
begin by establishing that there exists a simple spiraling path
$\xi$ on the surface of $O$ that starts and ends on the front rim of
the root band and winds around each band in $T_U$ at least once.
When this path is ``thickened'', it covers the band faces and
unfolds into a horizontal staircase-like strip to which front and
back faces of $O$ can be attached vertically. We describe $\xi$
recursively, starting with the base case in which $O$ consists of a
single box and thus the partition $\pi$ leaves a single band.

\subsection{Single Box Spiral Path}
\label{sec:basecase}

\begin{figure}[htbp]
\centering
\includegraphics[width=0.96\linewidth]{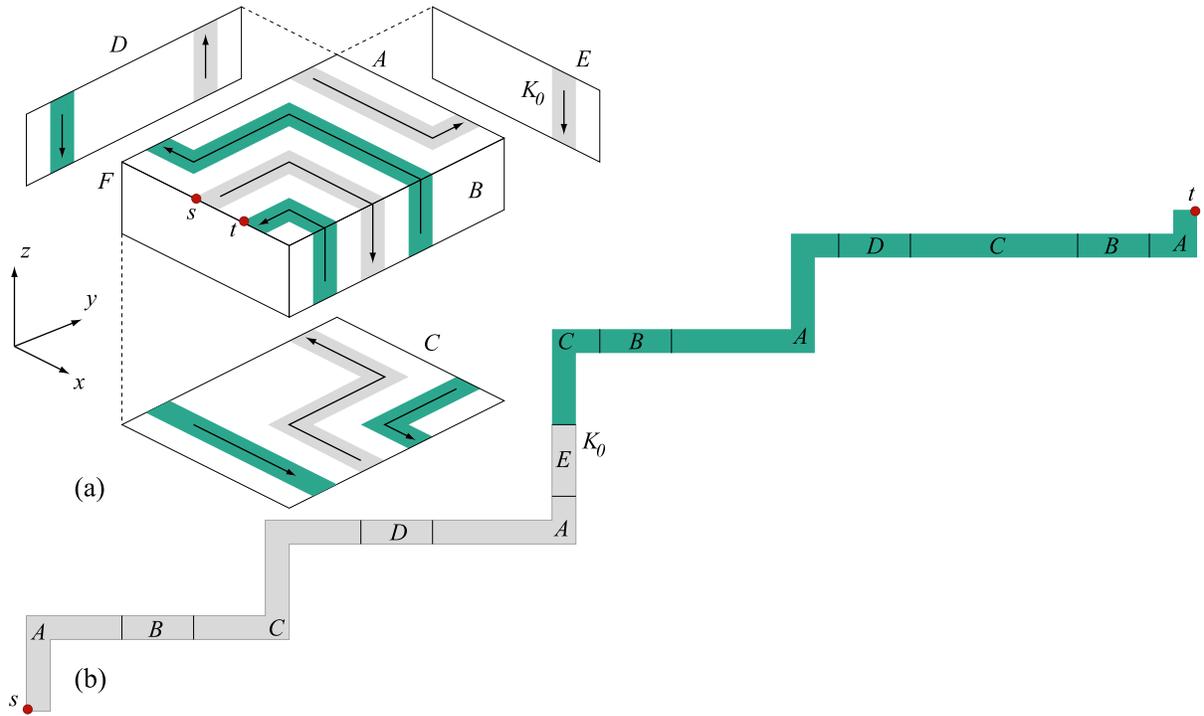}
\caption{(a) Spiral path $\xi$ with entering/exiting configuration
$R_{st}$: $s$ is left to $t$, and entering spiral heads rightward
from $s$. (b) Flattened spiral $\xi$.} \label{fig:box1}
\end{figure}

Let $O$ be a box with band $b$. We use the following notation (see
Figure~\ref{fig:box1}a): $A$,
$B$, $C$, and $D$ are top, right, bottom and left faces of $O$
(these faces belong to $b$); $E$ and $F$ are back and
front faces of $O$ (these faces do not belong to $b$); $s$ and
$t$ are {\em entering} and {\em exiting} points on the top edge of
the front rim of $b$.

The main idea is to start at $s$, spiral forward (front to back, cw
or ccw) around band faces $A$, $B$, $C$ and $D$, cross the back face
$E$ to reverse the direction of the spiral, then spiral backward
(back to front, ccw or cw) around band faces back to $t$. See
Figure~\ref{fig:box1}a for an example, where mirror views are
provided for the bottom, left, and back faces which cannot be viewed
directly. We refer to the forward spiral (incident to entering point
$s$) as the {\em entering} spiral, and the backward spiral (incident
to exiting point $t$) as the {\em exiting} spiral. Spiral $\xi$ is
the concatenation of the entering spiral, the back face strip, and
the exiting spiral. It can be unfolded flat and laid out
horizontally in a plane, as illustrated in Figure~\ref{fig:box1}b.

\begin{figure}[htbp]
\centering
\includegraphics[width=0.96\linewidth]{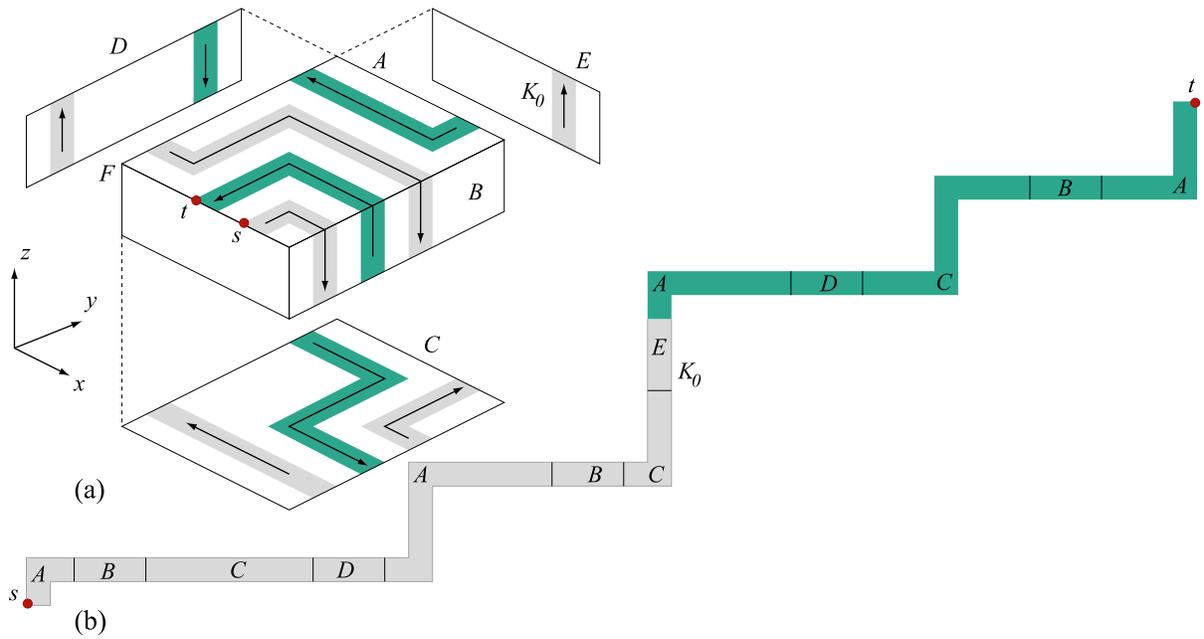}
\caption{(a) Spiral path $\xi$ with entering/exit configuration
$R_{ts}$: $s$ right of $t$; entering
spiral heads right from $s$. (b) Flattened spiral $\xi$.}
\label{fig:box2}
\end{figure}

\begin{figure}[htbp]
\centering
\begin{tabular}{c@{\hspace{0.05\linewidth}}c}
\includegraphics[width=0.45\linewidth]{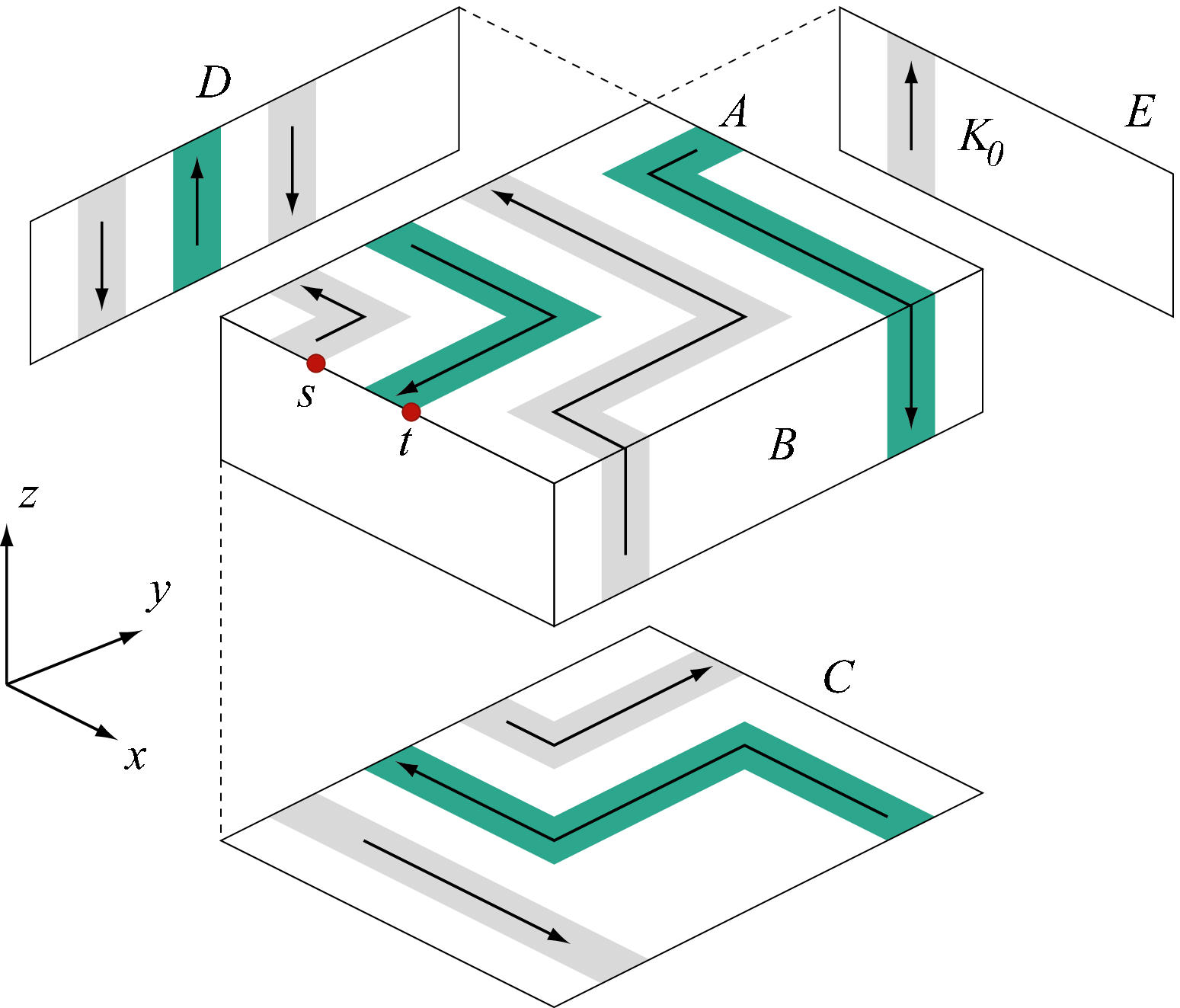} &
\includegraphics[width=0.45\linewidth]{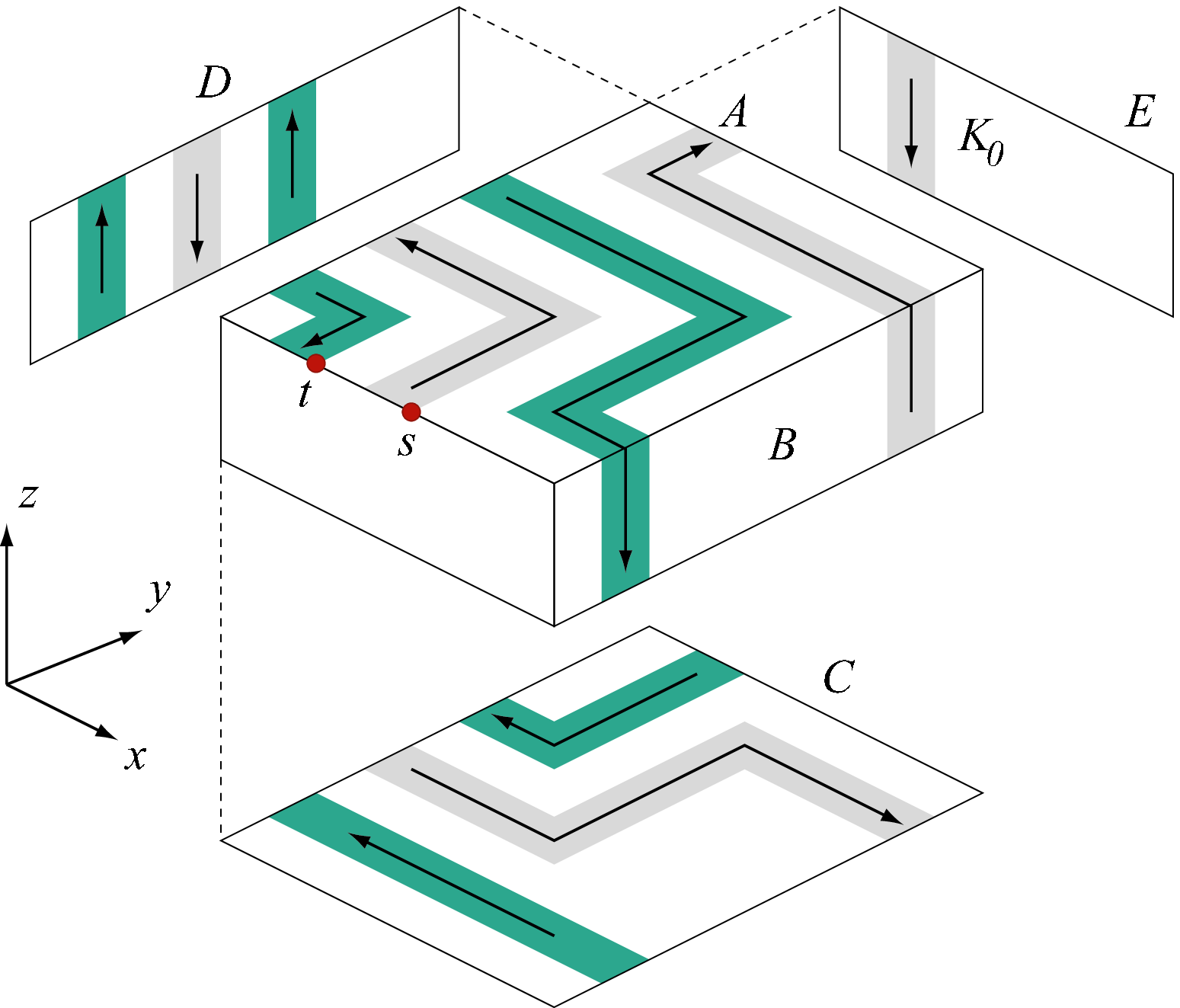} \\
(a) & (b)
\end{tabular}
\caption{(a) Spiral path $\xi$ with entering/exiting configuration
$L_{st}$: $s$ is left of $t$, and entering spiral heads left from
$s$. (b) Spiral path $\xi$ with entering/exiting configuration
$L_{ts}$: $s$ is right of $t$, and entering spiral heads left from
$s$.} \label{fig:box34}
\end{figure}

We distinguish four variations of this spiraling path, which differ
in how they enter and exit the band: the entering spiral is heading
away from $s$ either to the right (cw) or to the left (ccw), and $t$
is either to the left or to the right of $s$ on the front edge.
Directions left, right, cw, and ccw are defined from the perspective
of a viewer positioned at $y = -\infty$. We will use the notation
$R_{st}$, $R_{ts}$, $L_{st}$ and $L_{ts}$, to identify the four
possible entering/exiting configurations. Here the first letter ($R$
or $L$) indicates the direction ($R$ight or $L$eft) the entering
spiral is heading as it moves away from $s$, and the subscripts
indicate the position of $s$ relative to $t$. We use the symbol
$R\_$ for $b$ to indicate that the relative position of $s$ and $t$
is irrelevant to this discussion; thus $R\_$ denotes either $R_{st}$
or $R_{ts}$, and same for $L\_$. For the base case,
Figures~\ref{fig:box1}a, \ref{fig:box2}a, \ref{fig:box34}a
and~\ref{fig:box34}b illustrate all four configurations $R_{st}$,
$R_{ts}$, $L_{ts}$ and $L_{st}$, respectively. The unfoldings for
all four cases are similar, each flattening into a horizontal
staircase-like strip.

\begin{figure}[htbp]
\centering
\includegraphics[width=0.88\linewidth]{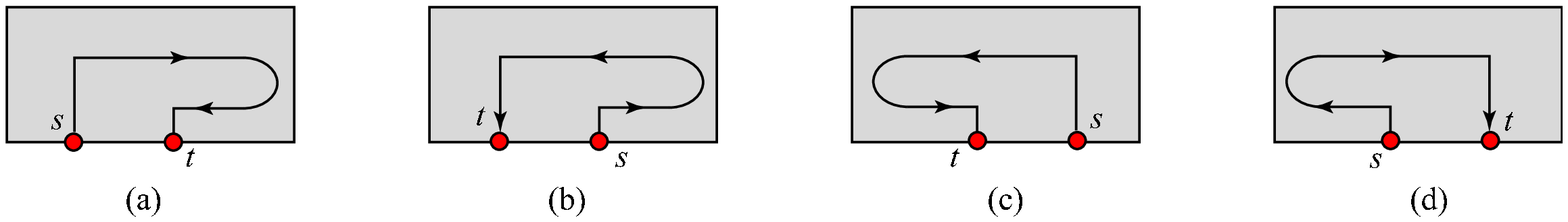}
\caption{2D representations of four base case variations of $\xi$
illustrating entering/exiting configurations (a) $R_{st}$ (b)
$R_{ts}$ (c) $L_{ts}$ (d) $L_{st}$.} \label{fig:rep.boxes}
\end{figure}

Three dimensional illustrations of $\xi$, such as 
Figures~\ref{fig:box1}a and~\ref{fig:box2}a, are impractical for all
but the smallest examples. To be able to illustrate more complex
unfoldings, we define a simple 2D representation for each of the
base case variations of $\xi$, as in Fig.~\ref{fig:rep.boxes}.
Note that each 2D representation captures 
the direction ($R$ or $L$) of the entering
spiral, and the relative position of $s$ and $t$. The entrance is
connected in a loop to the exit: the turnaround arc corresponds to
the forward spiral reversing its direction (cw to ccw, or ccw to cw)
using a back face strip (strip labeled $K_0$ in
Figures~\ref{fig:box1}a, \ref{fig:box2}a, ~\ref{fig:box34}).

\subsection{Recursive Structure}

In general, a band $b$ has children adjacent along its front and
back rims. The spiral path $\xi$ for the subtree rooted at $b$
begins and ends at two proximate points $s$ and $t$ on the top edge
of $b$'s front rim. The entering and exiting spirals of $b$ conform
to one of the four entering/exiting configurations $R_{st}$,
$R_{ts}$, $L_{st}$ or $L_{ts}$.

We describe $\xi$ at a high level first. Once $b$'s entering spiral
leaves $s$, it follows an alternating path to reach each of the
front children of $b$ and spiral around them recursively. An
alternating path is required because after spiraling around a child
of $b$, the direction of $b$'s spiral is reversed. After visiting
the front children, $b$'s entering spiral cycles forward around $b$
to its back rim, where it follows a second alternating path to reach
each of the back children and spiral around them recursively. After
visiting the last back child, $b$'s exiting spiral returns to $t$,
tracking the path taken by the entering spiral, but in reverse
direction. This final reverse spiral will revisit nodes/bands
already visited on the forward pass, and the recursive structure
will imply that some nodes/bands will be revisited many times before
the spiral returns to $t$.  We defer discussion of this consequence of
the algorithm to Section~\ref{sec:worst}.

\subsubsection{Alternating Paths for Labeling Children}
\label{sec:LR}

A preorder traversal of $T_U$ assigns each band an entering/exiting
configuration label ($R_{st}$, $R_{ts}$, $L_{st}$, or $L_{ts}$).
Although any label would serve for the root box of $T_U$, for
definitiveness we label it $R_{ts}$. We also pick an entering and
an exiting point on top of the root's front rim, with
the exiting point to the left of the entering point, which is consistent
with its $R_{ts}$ label.
In the following we provide
algorithms for labeling the front and back children of a labeled parent
$b$, which get applied when $b$ is visited during the traversal.
These rules are described in terms of two alternating paths that
$b$'s entering spiral takes to reach every front and back child. We
begin with the alternating path for labeling the front children.

\begin{figure}[htbp]
\centering
\includegraphics[width=0.85\linewidth]{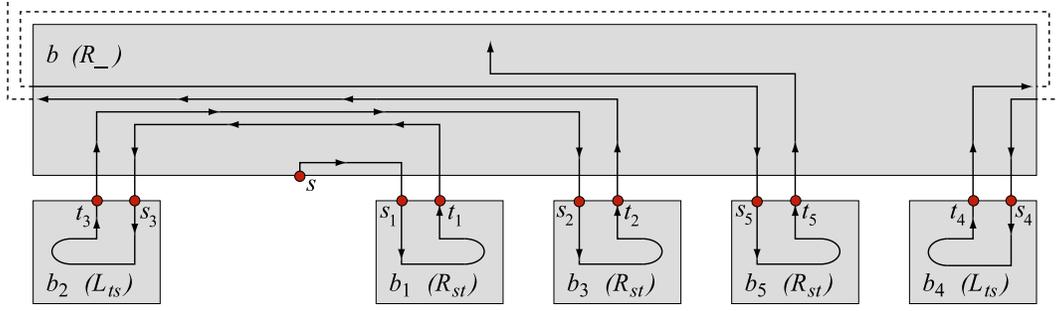}
\caption{Alternating path for labeling $b$'s front children.} \label{fig:label.front}
\end{figure}

\begin{center}\begin{minipage}{\linewidth}
\hrule\hfill \\
\noindent LABEL-FRONT-CHILDREN($b$) (see
Figure~\ref{fig:label.front}) \hrule\hfill
\begin{enumerate}
\item Set current position to $b$'s entering point $s$. Set current direction to
the direction of the entering spiral of $b$: rightward, if $b$ has
an $R$-label, and leftward if an $L$-label.
\item {\bf while} $b$ has unlabeled front children {\bf do}
   \begin{enumerate}
   \item From the current position, walk in the current direction
     along the front rim of $b$, until (i) an unlabeled front child $b_i$
     is encountered, and (ii) current position is on top of $b$.
   \item Assign to $b_i$ label $R_{st}$ if walking rightward, or $L_{ts}$
     if walking leftward.
     Select points $s_i$ and $t_i$ on the top line segment
     at the intersection between
     $b$ and $b_i$, in a relative position consistent with the label
     ($R_{st}$ or $L_{ts}$) of $b_i$.
     Reverse the current direction.
   \end{enumerate}
\end{enumerate}
\hrule\hfill
\end{minipage}\end{center}
We must reverse the current direction in Step~(b) above because
the spiral exits a child box heading in the
opposite direction from which it entered: if the entering spiral heads
rightward, the exiting spiral heads leftwards, and the other way
around. This forces the left/right alternation between the front
children of $b$.
Figure~\ref{fig:label.front} shows an example in which $b$ has five
front children and an $R$-type configuration. The children are
visited in the order $b_1$, $b_2$, $b_3$, $b_4$ and $b_5$; the
dashed lines correspond to walking around side and bottom faces of
$b$, to reach an unlabeled child from the top of $b$. The
configuration assigned to each child by the labeling procedure above
is shown within parentheses.

After labeling all front children of $b$, its back children are
labeled using a similar scheme. Unlike the situation for front
children, however, we have the flexibility of selecting which back
child to label first. For definitiveness, we always label first
either the leftmost or the rightmost back child, depending on
whether the alternating path heads leftward or rightward after
labeling the last front child. The procedure below describes the
alternating path used to label the back children.

\begin{figure}[htbp]
\centering
\includegraphics[width=0.85\linewidth]{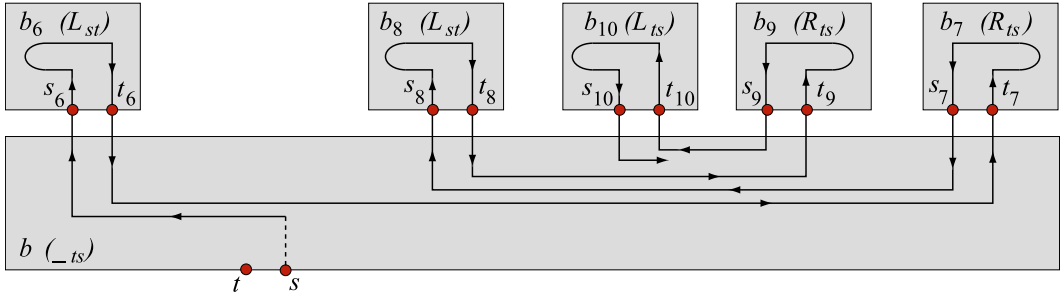}
\caption{Alternating path for labeling $b$'s back children.} \label{fig:label.back}
\end{figure}

\begin{center}\begin{minipage}{\linewidth}
\hrule\hfill \\
\noindent LABEL-BACK-CHILDREN($b$) (see Figure~\ref{fig:label.back})
\hrule\hfill
\begin{enumerate}
\item Set current position to $s$, if $b$ has no front children;
otherwise, set current position to the exiting point of the front
child last labeled.
\item Set current direction to the direction of $b$'s entering spiral,
if $b$ has no front children; otherwise, set current direction to
the direction of the exiting spiral of the front child of $b$
last labeled (i.e, leftward, if the child has an $R$-label, and
rightward if it has an $L$-label).
\item {\bf while} $b$ has unlabeled back children {\bf do}
  \begin{enumerate}
  \item If the current direction is leftward (rightward), then walk leftward
    (rightward) from the current position, until the leftmost
    (rightmost) unlabeled back child $b_i$ is encountered.
  \item If $b_i$ is not the last unlabeled back child, then assign
    to $b_i$ label $L_{st}$ ($R_{ts}$), if the current direction is
    leftward (rightward).
  \item If $b_i$ is the last unlabeled back child, assign to $b_i$ a
    label with the same ordering of $s$ and $t$ as for $b$. Specifically, if
    $b$ has a $\_{st}$ label and $b_i$ is entered
    while heading leftward (rightward), assign to $b_i$ label $L_{st}$
    ($R_{st}$); if $b$ has a $\_{ts}$ label and $b_i$ is
    entered while heading leftward (rightward), assign to $b_i$ label
    $L_{ts}$ ($R_{ts}$).
  \item Select points $s_i$ and $t_i$ on the top line segment
     at the intersection between
     $b$ and $b_i$, in a relative position consistent with the label of $b_i$.
     Reverse the current direction.
  \end{enumerate}
\end{enumerate}
\hrule\hfill
\end{minipage}\end{center}

Note that the relative position of $s_i$ and $t_i$ in the
entering/exiting configuration for the back child last labeled stays
consistent with the configuration for the parent.
Figure~\ref{fig:label.back} shows an example in which $b$ has five
back children, and a $\_ts$ unfolding configuration. The back
children get visited in the order $b_6, b_7, b_8, b_9, b_{10}$. The
unfolding label assigned to each back child is shown within
parentheses; observe that the unfolding label $\_ts$ for $b_{10}$
(the back child last labeled) is consistent with the unfolding label
$\_ts$ of its parent. Also observe that the nesting of the $L/R$
alternation is inside-out for front children, and outside-in for
back children (cf. Figures~\ref{fig:label.front}
and~\ref{fig:label.back}).

The path exiting $b_{10}$ must now return to the exiting point $t$
of parent $b$, and to do so, because it is deeply nested in
the alternating paths, it must follow the entire path between
the entering point of $b$ and the entering point of $b_{10}$, but in
reverse direction. We return to this in the next section.

\subsubsection{Recursive Spiral Paths}
\label{sec:recursive.spiral}
\begin{lemma}
For any unfolding tree $T_U$ rooted at a node with entering point
$s$ and exiting point $t$, there exists a simple spiral path $\xi$
such that (i) $\xi$ starts at $s$, cycles around each band in $T_U$
at least once, and returns to $t$ heading in the reverse direction
(ii) for each
node $b$ in $T_U$, $\xi$ is consistent with the entering/exiting
configuration for $b$, and (iii) $\xi$ unfolds flat horizontally,
with $s$ on the far left and $t$ on the far right.
\label{lem:xi.exist}
\end{lemma}
\begin{proof}
The proof is by induction on the depth of $T_U$. The base case
corresponds to a tree with a single node $b$ (of depth $1$),
and is established by
Figures~\ref{fig:box1}a,~\ref{fig:box2},~\ref{fig:box34}a, and~\ref{fig:box34}b.

Assume that the lemma holds for any unfolding tree of depth $d$ or
less, and consider an unfolding tree $T_U$ of depth $d+1$ rooted at
$b$. We only discuss the case in which $b$ has an $R_{ts}$
entering/exiting configuration; the other three cases ($R_{st}$,
$L_{ts}$ and $L_{st}$) are similar.

\begin{figure}[htbp]
\centering
\includegraphics[width=0.75\linewidth]{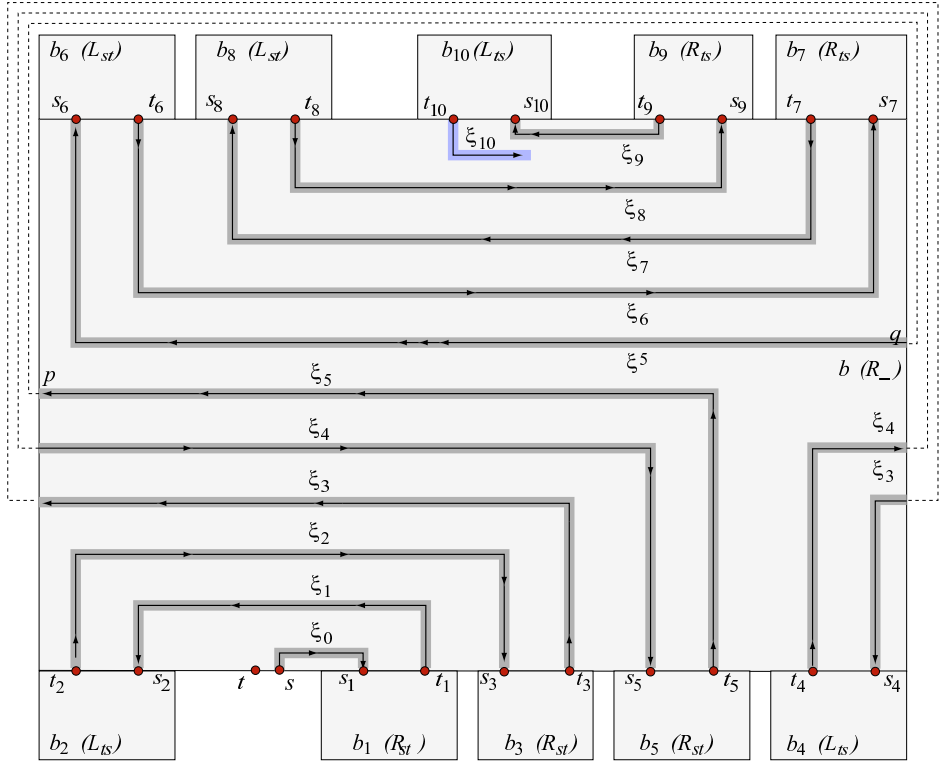}
\caption{Strip segments along the
front and back alternating paths used to reach the children of $b$:
$\xi_0$ is used to get from $s$ to $b_1$, $\xi_1$ from $b_1$ to
$b_2$, and so on.} 
\label{fig:xi.enter}
\end{figure}

We begin with the general case when $b$ has both front and back
children. Let $b_1, b_2 \ldots b_k$ be the front children and
$b_{k+1}, b_{k+2} \ldots$ be the back children of $b$, in the order in
which they are visited by the labeling procedures from
section~\ref{sec:LR}. The spiral $\xi$ starts at $s$ and follows the
front alternating path illustrated in Figure~\ref{fig:label.front}
to reach each front child $b_i$. We apply the inductive hypothesis
on $b_i$ to conclude the existence of a spiral path $\xi(b_i)$ from
$s_i$ to $t_i$ for the subtree rooted at $b_i$. From $t_i$, $\xi$
continues along the front alternating path to the next front child.
Figure~\ref{fig:xi.enter} illustrates the strip segments along the
front and back alternating paths used to reach the children of $b$:
$\xi_0$ is used to get from $s$ to $b_1$, $\xi_1$ from $b_1$ to
$b_2$, and so on.

After visiting the last front child $b_k$, $\xi$ cycles around $b$
once, stopping at the entering point $s_{k+1}$ of $b_{k+1}$ (see
spiral segment $\xi_5$ in Figure~\ref{fig:xi.enter}). This cycle is
necessary to ensure that $\xi$ goes around $b$ at least once. From
$s_{k+1}$, spiral $\xi(b_{k+1})$ takes $\xi$ to $t_{k+1}$, and from
there $\xi$ moves along the back alternating path on to the next
back child. The portion of $\xi$ from $s$ to the entering point of
the back child last visited is $b$'s entering spiral. In the example
of Figure~\ref{fig:xi.enter}, the entering spiral starts at $s$ and
ends at $s_{10}$.

\begin{figure}[htbp]
\centering
\includegraphics[width=\linewidth]{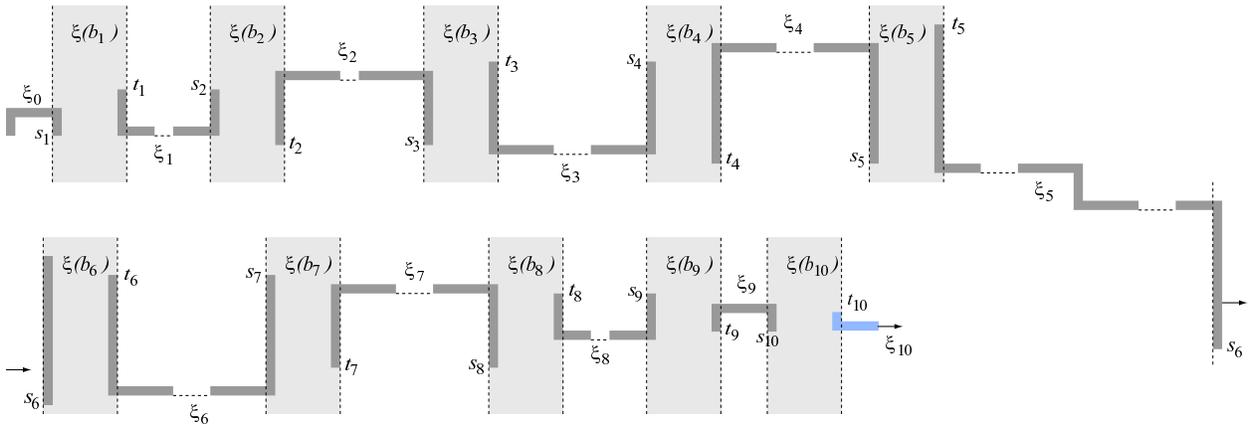}
\caption{Band $b$'s entering spiral unfolded. 
} \label{fig:xi.enter.unf}
\end{figure}

The unfolding of $b$'s entering spiral is depicted in
Figure~\ref{fig:xi.enter.unf}, where each shaded region contains the
horizontal unfolding of $\xi(b_i)$ corresponding to the subtree
rooted at $b_i$. Linking the children's spirals together are the
strip segments of the alternating paths and the cycle around $b$.

\begin{figure}[htbp]
\centering
\includegraphics[width=0.75\linewidth]{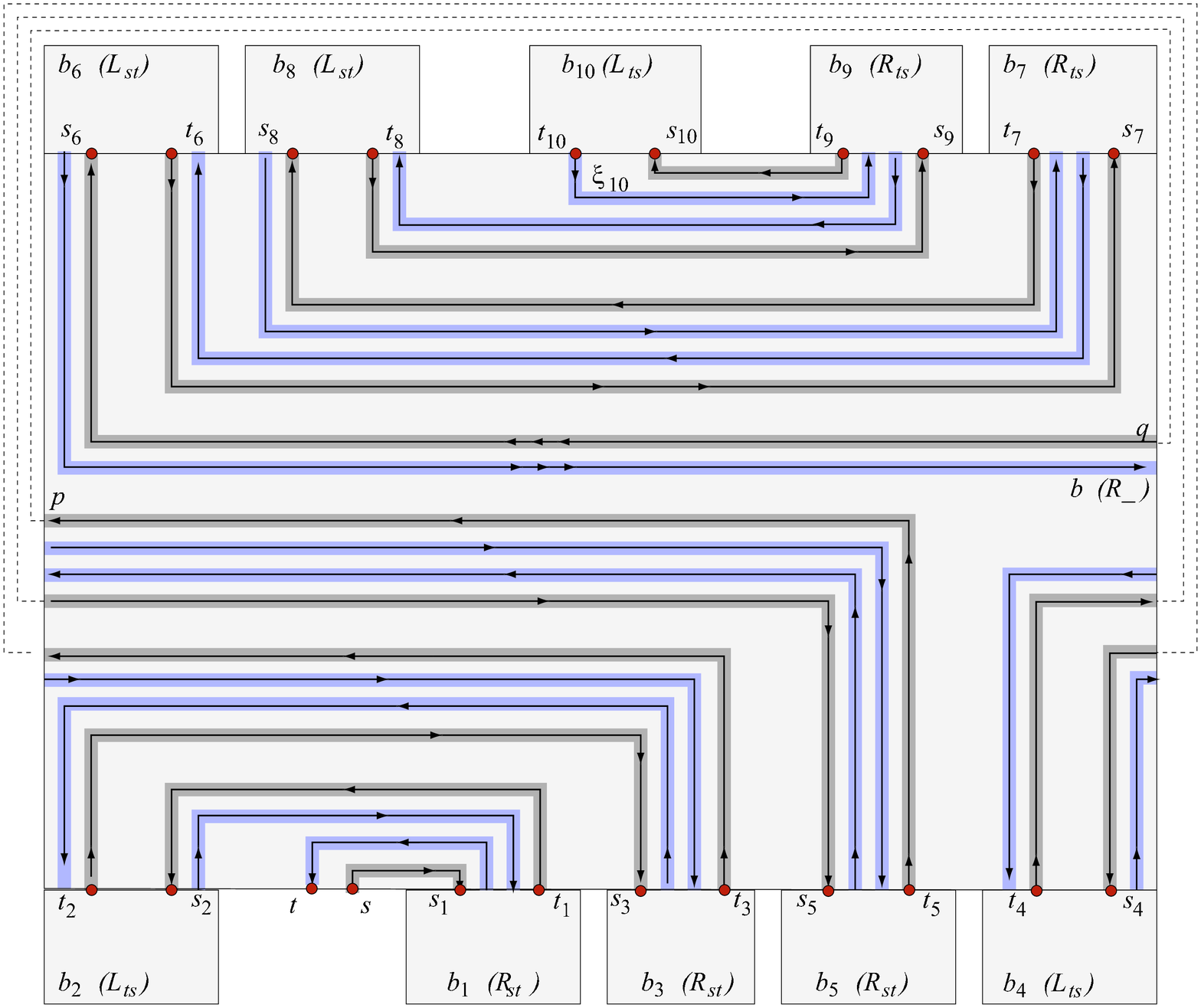}
\caption{Forward and backward $\xi = \xi(b)$.} \label{fig:xi.exit}
\end{figure}

Once $\xi$ leaves the last back child, it must return to the exiting
point $t$ of $b$, and it does so by tracking the entering spiral of
$b$ in the reverse direction. This portion of $\xi$ is $b$'s exiting
spiral. By making the entering spiral arbitrarily thin and
positioning it so that it doesn't touch any edge of $O$ or itself,
we ensure that there is room along its sides for the exiting spiral.
Figure~\ref{fig:xi.exit} illustrates all fragments of $\xi$ that
belong to $b$.

\begin{figure}[hptb]
\centering
\includegraphics[width=0.85\linewidth]{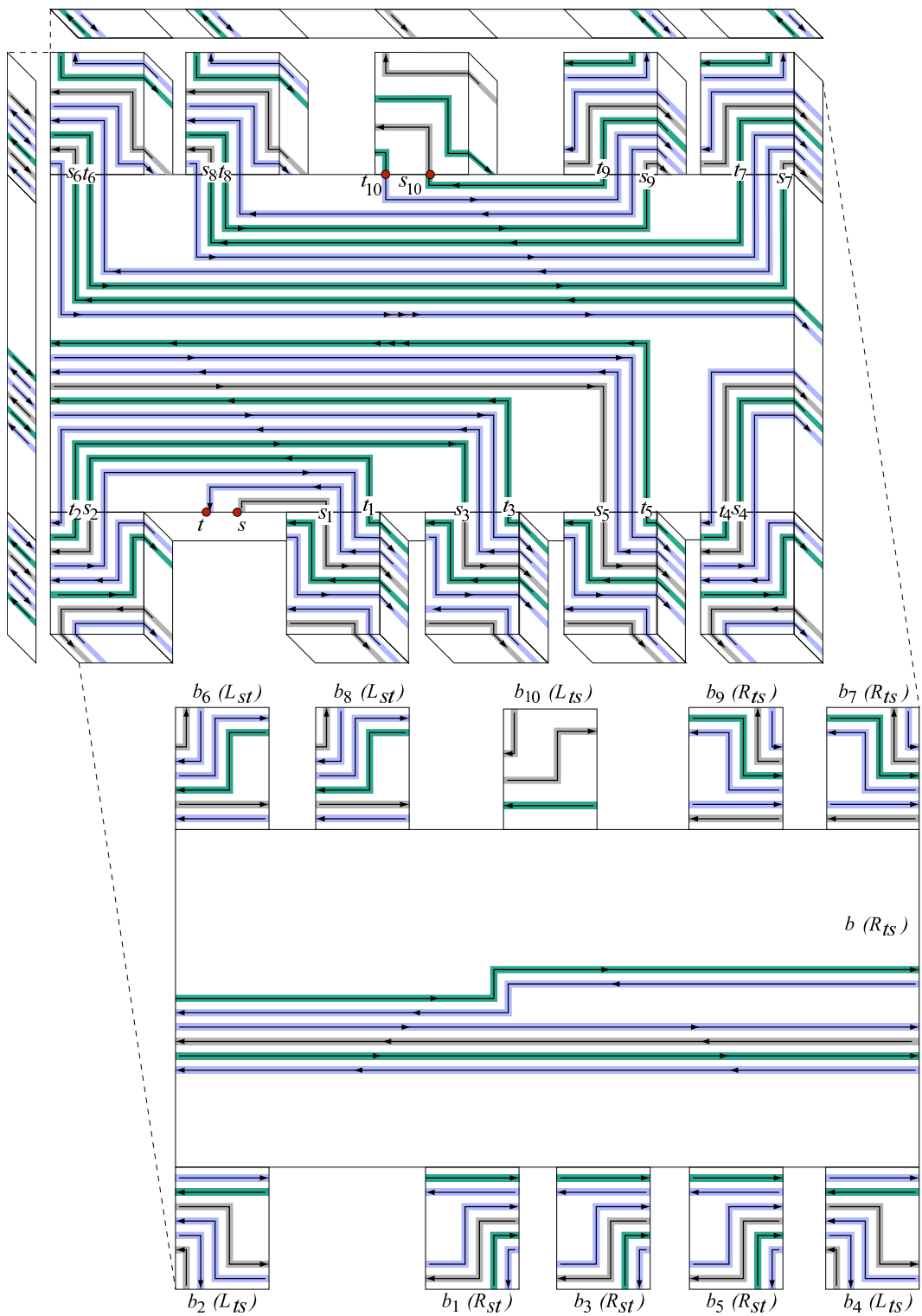}
\caption{Spiral $\xi$ for a complete example, with mirror views for
left, back and bottom faces. Entering spiral for $b$ extends from
$s$ to $s_{10}$ and exiting spiral from $t_{10}$ back to $t$.}
\label{fig:xi.3d}
\end{figure}
As a complete running example, Figure~\ref{fig:xi.3d} illustrates
the spiral $\xi$ in its entirety for the case in which none of the
children $b_i$ has children of its own. The spiral for each $b_i$
corresponds to one of the base cases (see
Figs.~\ref{fig:box1},~\ref{fig:box2} and~\ref{fig:box34}). The
entering spiral of $b$ extends from $s$ to $s_{10}$ and visits $b_1,
b_2, \ldots, b_{10}$ in this order. The exiting spiral of $b$
extends from $t_{10}$ to $t$ and visits $b_9, b_8 \ldots b_1$ in
this order, tracking closely the entering spiral in reverse.

We now prove that $\xi$ satisfies the three conditions stated in the
lemma. It is clear that $\xi$ cycles around each band in $T_U$ at
least once: by induction, $\xi(b_i)$ cycles around each band in the
subtree rooted at $b_i$ at least once, and $\xi$ cycles around $b$
after visiting the front children. Also note that the exiting spiral
ends up at $t$, left of $s$. This is because the last
visited back child and $b$ have the same $\_st$ or $\_ts$ label,
and $b$'s exiting spiral tracks its entering spiral in reverse
from the last back child's exiting point to $t$.
In Figure~\ref{fig:xi.enter},
the last back child $b_{10}$ has $t_{10}$ to the left of $s_{10}$,
which places $b$'s entering spiral to the left of its exiting
spiral, guaranteeing that the exiting spiral terminates to the left of
$s$. Therefore, $\xi$ satisfies condition (i) of the lemma. By
induction, $\xi(b_i)$ is consistent with every configuration in the
subtree rooted at $b_i$, therefore $\xi$ satisfies condition (ii) of
the lemma as well.

Figure~\ref{fig:xi.enter.unf} shows the horizontal unfolding of the
entering spiral (from $s$ to $s_{10}$). The unfolding of the exiting
spiral is similar, but rotated $180^\circ$. The unfolding of the
entire spiral $\xi$ is the concatenation of the unfolded entering
spiral, the last back child's spiral, and the
exiting spiral. It can be easily verified that this satisfies
condition (iii) of the lemma. This completes the proof for the case
when $b$ has front and back children.

If $b$ has no front children, then from $s$ the spiral proceeds to
cycling around $b$ and then alternating between the back children.
If there are no back children, then after visiting the front
children and cycling around $b$, the spiral reverses its direction
using a strip from the back face of $b$ (as done in the base case),
then tracks the entering spiral back to $t$.
\end{proof}

\medskip
\noindent The proof of Lemma~\ref{lem:xi.exist} leads to an
algorithm for computing the spiral path for a subtree rooted at band
$b$, as described in the procedure SPIRAL-PATH($b$) below.
\begin{center}\begin{minipage}{\linewidth}
\hrule\hfill \\
\noindent SPIRAL-PATH($b$) \hrule\hfill
\begin{enumerate}
\squeezelist
\item If $b$ has no children, follow the appropriate base-case
spiral path and return.
\item If $b$ has no front children, skip to Step 3.
\item {\bf while} ($b$ has unvisited front children) {\bf do}
\begin{enumerate}
\item [2.1] Follow the front alternating path to
the entering point of next front child $b_i$ of $b$.
\item [2.2] SPIRAL-PATH($b_i$).
\end{enumerate}
\item Complete a cycle around $b$ and proceed to the back rim of
$b$.
\item If $b$ has no back children, reverse spiral using a back face
strip and skip to Step 6.
\item {\bf while} ($b$ has unvisited back children) {\bf do}
\begin{enumerate}
\item [4.1] Follow the back alternating path to the
entering point of next back child $b_i$ of $b$.
\item [4.2] SPIRAL-PATH($b_i$).
\end{enumerate}
\item Retrace the entering spiral for $b$ back to the exiting point of $b$.
\end{enumerate}
\hrule\hfill
\end{minipage}\end{center}

\subsection{Recursive Spiral Path Example}

\begin{figure}[htbp]
\centering
\includegraphics[width=0.73\linewidth]{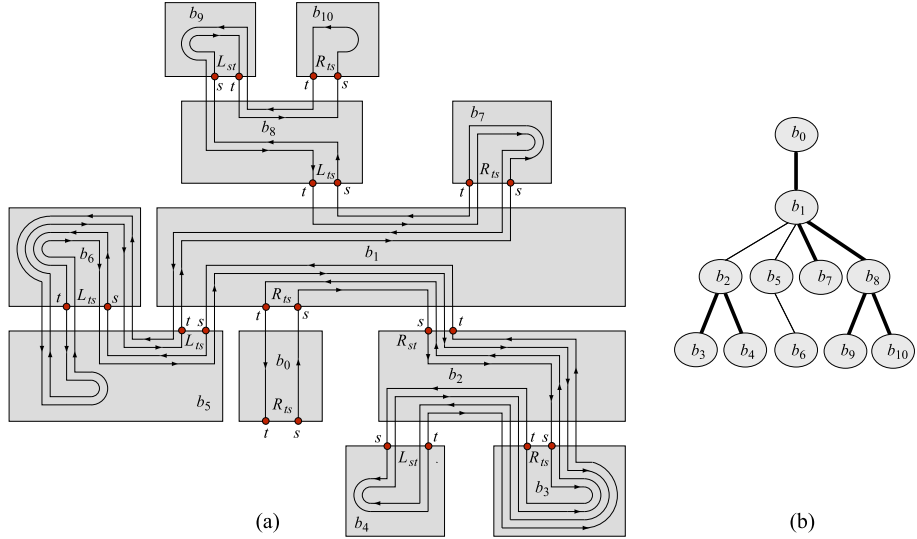}
\caption{A complete recursive unfolding example: 2D representation.}
\label{fig:ex.robin}
\end{figure}

To reinforce our recursive spiraling ideas, we provide the 2D
representation of an unfolding for a slightly more complex example,
illustrated in Fig.~\ref{fig:ex.robin}. Observe that the cycle that
$\xi$ makes around each band is not captured by a 2D representation,
therefore we omit mentioning it in this section.

We first describe the structure of the unfolding tree for our
example. Box $b_0$ is the root of the unfolding tree; it has one
back child $b_1$ and no front children. Box $b_1$ has two front
children, $b_2$ and $b_5$, and two back children, $b_7$ and $b_8$.
Box $b_2$ has two back children, $b_3$ and $b_4$, and no front
children. Box $b_5$ has one front child $b_6$ and no back
children. Box $b_8$ has two back children, $b_9$ and $b_{10}$, and
no front children.

The algorithm begins by assigning an $R_{ts}$ label to $b_0$, and then
uses the procedure from Section~\ref{sec:LR} to assign labels to
the other boxes, as marked on each box in Fig.~\ref{fig:ex.robin}.

We now discuss the order in which the spiral $\xi$ visits these
boxes. From $b_0$, $\xi$ enters the back child $b_1$ and then
proceeds along the front alternating path to reach the front child
$b_2$ first. Since $b_2$ has no front children, $\xi$ proceeds to
the back of $b_2$ to visit its back children $b_3$ and $b_4$, in
this order. Once it exits the last back child $b_4$, $\xi$ begins
tracking the entering spiral for $b_2$ in reverse back to reach $t$
on the front rim of $b_2$, thus visiting $b_3$ again. From the front
rim of $b_2$, $\xi$ follows $b_1$'s front alternating path to $b_5$,
then immediately to front child $b_6$ and then to the back of $b_5$.
Since $b_5$ has no back children, $\xi$ reverses direction using a
back face strip and begins tracking the path back to $b_1$, thus
visiting $b_6$ again. Note that $\xi$ is moving rightwards as it
reenters $b_1$, therefore it proceeds to the rightmost back child
$b_7$ of $b_1$. Upon exiting $b_7$, $\xi$ moves along $b_1$'s back
alternating path to $b_8$. Note that $b_8$ is the last back child of
$b_1$ to be visited, therefore the spiral segment $\xi^*$ between
the entering point of $b_1$ and the entering point of $b_{8}$ is the
entering spiral of $b_1$. Also note that $\xi^*$ visited boxes in
the order $b_1, b_2, b_3, b_4, b_3, b_2, b_1, b_5, b_6, b_5, b_6,
b_5, b_1, b_7, b_1$, therefore the exiting spiral of $b_1$ will
revisit the same sequence in reverse order on its way back to the
front rim of $b_1$. Between the entering and exiting points of
$b_8$, $\xi$ visits the back children $b_9$ and $b_{10}$, then $b_9$
again on its way back to the front of $b_8$. Fig.\ref{fig:ex.robin}
shows $\xi$ in its entirety.

\subsection{Thickening $\xi$}
\label{sec:thicken}

\begin{figure}[htbp]
\centering
\includegraphics[width=0.85\linewidth]{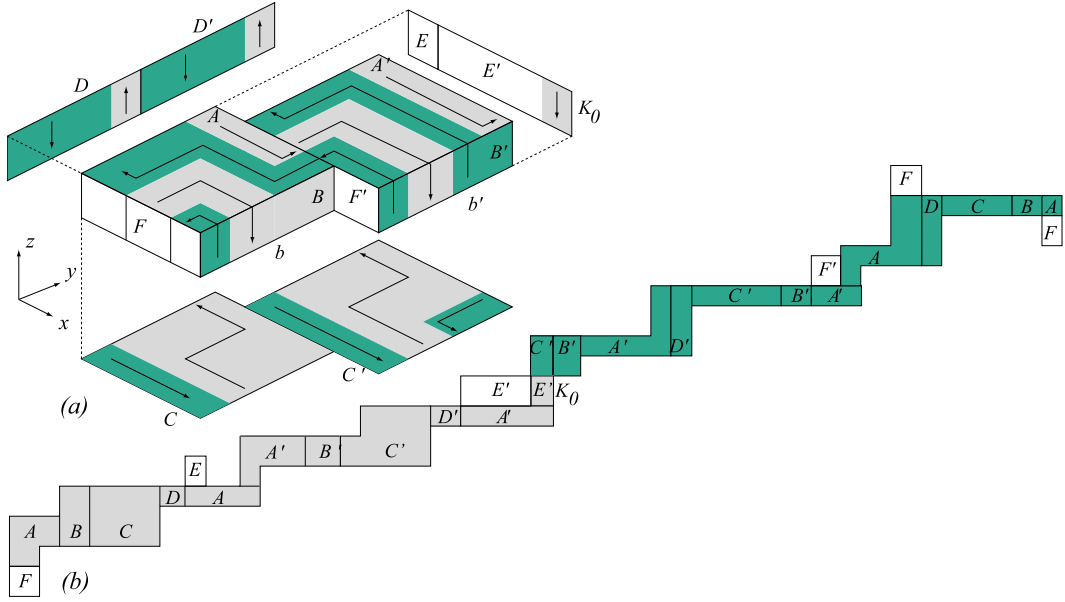}
\caption{(a) Thickened $\xi$ entirely covers $b$ and $b'$. (b) Planar layout of
spiral $\xi$ with front and back face pieces attached above and below.} 
\label{fig:boxfill1}
\end{figure}

Spiral $\xi$ established in Section~\ref{sec:recursive.spiral} can
be thickened in the $y$ direction so that it entirely covers each
band. This results in a vertically thicker unfolded strip. See
Figure~\ref{fig:boxfill1} for an example that illustrates the
thickening procedure on the base case from Figure~\ref{fig:box1}.
Since the unfolded $\xi$ is monotonic in the horizontal direction,
thickening it vertically cannot result in overlap. From this point
on, whenever we refer to $\xi$, we mean the thickened $\xi$.

\subsection{Attaching Front and Back Faces}
\label{sec:front.back}

Finally, we ``hang'' the front and back faces of $O$ from $\xi$ in 
a manner similar to that done in~\cite{dfo-gvuop-06}, as follows.
Consider the set of top edges of $O$ that separate
band faces from front or back faces. These edges are each part of a
rim, and hence they are found on the horizontal boundaries of the
unfolded $\xi$ as a collection of one or more contiguous segments.
We partition the front and back faces of each band $b$ by imagining
the top edges on the rim of $b$ illuminating downward lightrays in
these faces. This illuminates all front and back pieces;
these pieces are attached above and below $\xi$ to the corresponding
illuminating rim segments. 

For an example, see Figure~\ref{fig:boxfill1} showing a two band shape
and its unfolding. Here the front face of band $b$ is partitioned into
three pieces which are hung from their corresponding rim segments in
the unfolding. Faces $E$, $F'$ and $E'$ are hung similarly. Observe
that an example of edge overlap mentioned in Section~\ref{sec:Intro} occurs
between face $F'$ and and a section of $A$ in the unfolding.

This completes the unfolding process, which we summarize in the
procedure UNFOLD-EXTRUSION($O$) below.

\begin{center}\begin{minipage}{\linewidth}
\hrule\hfill \\
\noindent UNFOLD-EXTRUSION($O$) \hrule\hfill
\begin{enumerate}
\squeezelist
\item Partition $O$ into bands with $xz$ parallel planes $Y_0, Y_1, \ldots$
through each vertex (Section~\ref{sec:Intro}).
\item Compute unfolding tree $T_U$ with root band $b_0$.
\item Select root band $b_0$ adjacent to $Y_0$ and compute unfolding tree $T_U$
with root $b_0$.
\item For each band $b$ encountered in a preorder traversal of $T_U$
\begin{itemize}
\item[3.1] LABEL-FRONT-CHILDREN($b$).
\item[3.2] LABEL-BACK-CHILDREN($b$) (Section~\ref{sec:LR}).
\end{itemize}
\item Determine $\xi$ = SPIRAL-PATH($b_0$) (Sections~\ref{sec:basecase},~\ref{sec:recursive.spiral}).
\item Thicken $\xi$ to cover all bands in $T_U$ (Section~\ref{sec:thicken}).
\item Hang front and back faces off $\xi$ (Section~\ref{sec:front.back}).
\end{enumerate}
\hrule\hfill
\end{minipage}\end{center}

\section{Unfolding All Orthogonal Polyhedra}

The unfolding algorithm described for extrusions generalizes to
unfolding all orthogonal polyhedra. Let $O$ be a genus-zero
orthogonal polyhedron.  The surface of $O$ is simply connected,
which means that any closed curve on the surface can be continuously
contracted on the surface to a point.  We will use this
characterization in our proofs. We start by partitioning $O$ into
bands with $xz$ parallel planes $Y_0, Y_1,...,Y_i,...$ through each
vertex.

\subsection{Determining Connecting $z$-Beams}
\label{sec:connecting.beams}

Define a {\em z-beam} to be a vertical rectangle on the surface of
$O$ of nonzero width connecting two band rims.  
In the degenerate case, a $z$-beam has height zero
and connects two rims along a section where they coincide. 
We say that two
bands $b_1$ and $b_2$ are {\em z-visible} if there exists a $z$-beam
connecting an edge of $b_1$ to an edge of $b_2$.

\begin{figure}[htbp]
\centering
\includegraphics[width=0.4\linewidth]{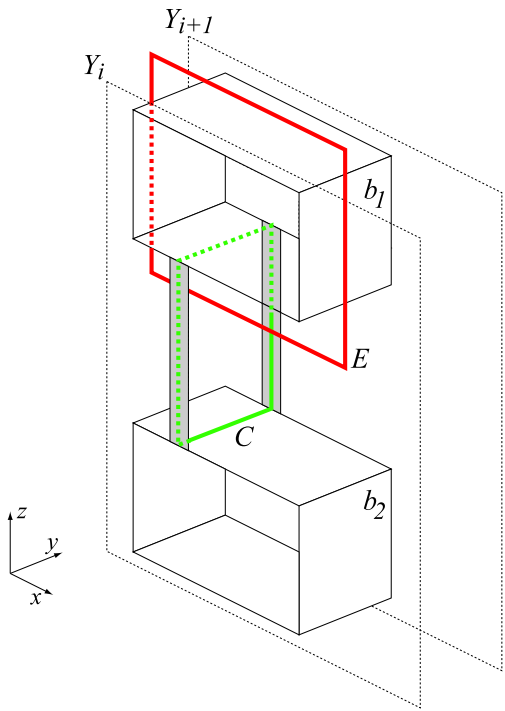}
\caption{If two $z$-visible bands are connected by beams to both
rims of each, then the surface curve $C$ is interlinked with
exterior curve $E$.} \label{fig:two.zbeams}
\end{figure}

\begin{lemma}
All $z$-beams between two $z$-visible bands lie in one $Y_i$ plane.
\label{lem:two.zbeams}
\end{lemma}
\begin{proof}
Suppose to the contrary that bands $b_1$ and $b_2$ are connected by
beams in both $Y_i$ and $Y_{i+1}$, i.e., both rims of both bands are
connected by $z$-beams. Then we can construct a closed curve $C$ on
the surface of $O$ from $b_1$, following the beam on $Y_i$ to $b_2$,
and following the beam on $Y_{i+1}$ back to $b_1$. See
Figure~\ref{fig:two.zbeams}. Now let $E$ be a closed curve just
exterior to, say $b_1$, parallel to and between $Y_i$ and $Y_{i+1}$.
Then $E$ and $C$ are interlinked. This means that $C$ cannot be
contracted to a point, contradicting the genus-zero assumption. 
\end{proof}

Thus all the $z$-beams between two $z$-visible bands are in this
sense equivalent.  We select one $z$-beam of minimal (vertical) length
to represent this equivalence class.

\subsection{Computing Unfolding Tree $T_U$}
\label{sec:unf.tree}

Let $G$ be the graph that contains a node for each band of $O$ and
an arc for each pair of $z$-visible bands. It easily follows from
the connectedness of the surface of $O$ that $G$ is connected.
Let the unfolding tree $T_U$ be any spanning tree of $G$, with the
root selected arbitrarily from among all bands adjacent to $Y_0$.

As defined in Section~\label{sec:Introduction},, the rim of the root node/band at
$y_0$ is called its \emph{front} rim; the other is its {\em back}
rim.  For any other band $b$, we provide definitions equivalent to
the ones in Section~\ref{sec:unfolding.extrusions}, only this time
in terms of connecting $z$-beams: the {\em front} rim of $b$ is the
one to which the (representative) $z$-beam to its parent is
attached; and the other rim of $b$ is its {\em back} rim
(Lemma~\ref{lem:two.zbeams} guarantees that this definition is
unambiguous.)  A child is a {\em front child} ({\em back child}) if
its $z$-beam connects to the front (back) rim of its parent.  We
call the region of the $Y$-plane enclosed by a band's back rim its
{\em back face}, and we say that the back face is {\em exposed} if
it is a face of $O$.

\begin{figure}[htbp]
\centering
\includegraphics[width=0.55\linewidth]{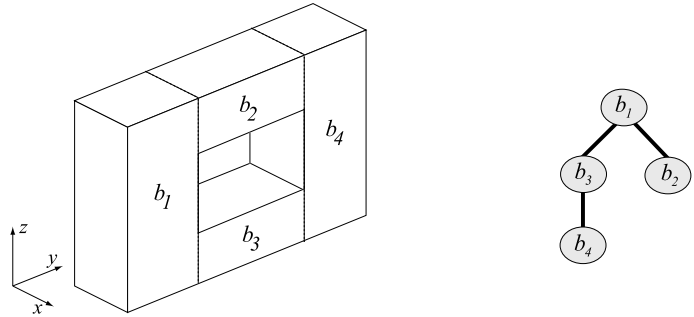}
\caption{For this genus-$1$ polyhedron, the spanning tree on the
right contains a band ($b_2$) with no back children and an unexposed
back face.} \label{fig:genus1}
\end{figure}

The following lemma establishes that bands with no back children in
$T_U$ have exposed back faces. This will be important in proving the
correctness of our unfolding algorithm, because we will need to
employ strips from the exposed back faces to turn the spiral around,
analogous to the $K_0$ strip in Figure~\ref{fig:box1}.
We note that this lemma is not true if $O$ has a non-zero genus.
Figure~\ref{fig:genus1}, for example, shows a polyhedron with a hole
in it, whose corresponding spanning tree (depicted on the right)
contains a band ($b_2$) with no back children and an unexposed back
face.

\begin{lemma}
The back face of every band in $T_U$ with no back children is
exposed. \label{lem:back.exposed}
\end{lemma}
\begin{proof}
We begin by establishing that any two band points $p$ and $q$ of $O$
are connected by a simple surface curve that follows the path in $T_U$
between $p$'s band and $q$'s band. Let
$(b_1, b_2, ..., b_k)$ be the path in $T_U$ between band $b_1$
containing $p$ and band $b_k$ containing $q$, and let $z_1, z_2, ...,
z_{k-1}$ be the $z$-beams connecting pairs of adjacent nodes along
this path.  From $p$, the surface curve moves to an arbitrary position
on the rim of $b_1$, then around the rim until it meets $z_1$,
then along $z_1$ to the rim of $b_2$. In a similar manner the
curve moves from $b_2$ to $b_3$ and so on, until it reaches
$b_k$ (more precisely, the line segment at the intersection
between $z_{k-1}$ and $b_k$).  Once on $b_k$, the curve moves
along the rim of $b_k$ up to the point $y$-opposite to $q$, and
finally in the $y$-direction to $q$. Figure~\ref{fig:simplecurve}
shows the surface curve corresponding to a five band path.
%
\begin{figure}[htbp]
\centering
\includegraphics[width=0.6\linewidth]{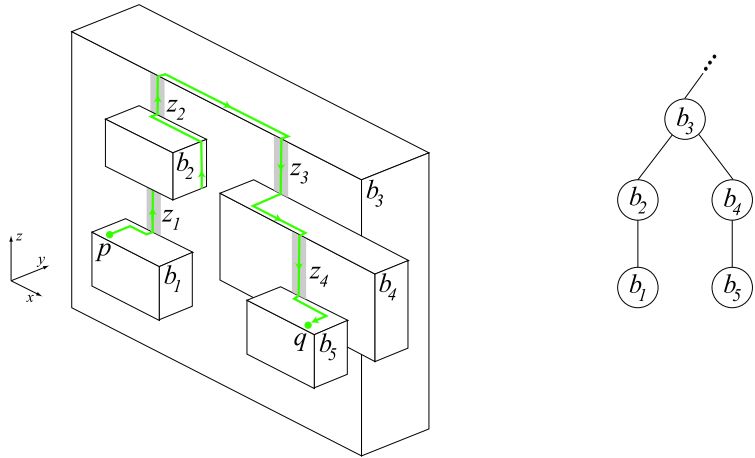}
\caption{A surface curve corresponding to a five-band path in $T_U$
(on the right).} \label{fig:simplecurve}
\end{figure}

Now suppose for the sake of contradiction that there is a band $b$
with no back children whose back face is not exposed. Let $r_-$ be
the back rim of $b$ and $r_+$ the front rim of $b$. The rim $r_-$
connects to the surface of $O$, 
so from any point on $r_-$, we can shoot a vertical ray on $O$'s surface 
and it will hit another rim point, either another point on
$r_-$ or a point on some other rim. (Generally this ray will
lie in a front or back face of $O$ adjacent to $r_-$, but in the
degenerate case when $r_-$ coincides with another rim, the height of
the ray will be zero).
If the vertical rays from $r_-$ 
only hit $r_-$ again, then in fact
the back face of $b$ is exposed. So there must be some vertical ray
$\alpha$ that extends from $r_-$ to a point $p'$ on some other band
$b'$. See Figure~\ref{fig:zerobackchildren}.

%
\begin{figure}[htbp]
\centering
\includegraphics[width=0.32\linewidth]{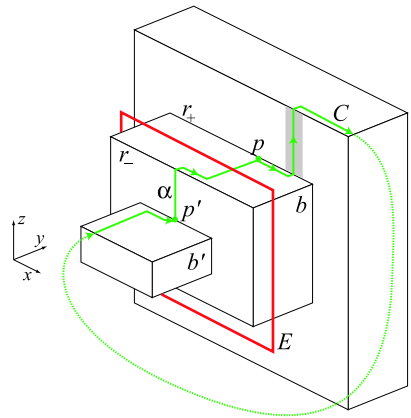}
\caption{For a band $b$ with no back children in $T_U$, an unexposed
back face implies two interlinked closed curves, $C$ on the surface
of $O$ and $E$ exterior.} \label{fig:zerobackchildren}
\end{figure}

Let $p$ be a point on the front rim $r_+$ of $b$. We established
above the existence of a particular surface curve $C$ from $p$ to
$p'$ corresponding to the path between $b$ and $b'$ in $T_U$.
Because $b$ has no back children in $T_U$, $C$ moves from $r_+$ to
the parent of $b$ and never returns to $b$ again, therefore it never
touches $r_-$ (refer to Figure~\ref{fig:zerobackchildren}).
We now extend $C$ to a simple closed curve that crosses $r_-$ as
follows: from $p'$, $C$ travels along the vertical ray $\alpha$ to
$r_-$, then around $r_-$ until it reaches the point $y$-opposite
to $p$, and finally it extends in the $y$-direction to $p$. Now
consider a second closed curve $E$ exterior to $O$ that cycles
around band $b$. Curves $C$ and $E$ are interlinked, meaning
that $C$ cannot be contracted to a point. This contradicts the
genus-zero assumption.
\end{proof}

\subsection{Unfolding Algorithm}

The algorithm that unfolds all bands in $T_U$ is very similar to the
algorithm that unfolds extrusions (described in
Section~\ref{sec:unfolding.extrusions}). The main difference is that
spiral $\xi$ must now travel along the vertical $z$-beams that connect
a parent to its children; these $z$-beams unfold vertically in the
plane. The unfolding starts by assigning an entering/exiting
configuration to each band in $T_U$ (as in Section~\ref{sec:LR}), then
determines a spiral $\xi$ with the properties listed in
Lemma~\ref{lem:xi.exist}. Recall that $\xi$ follows the (front, back)
alternating paths on every band $b$ in $T_U$, in order to reach all
(front, back) children of $b$. Unlike for extrusions however, where an
alternating path leads directly to a child $b_i$ of $b$, in this case
such an alternating path leads to a $z$-beam $\alpha$ connecting $b_i$
to $b$; therefore, $\xi$ must continue along $\alpha$ to reach $b_i$.

For bands with no back children, $\xi$ reverses direction using a
strip from its back face. Lemma~\ref{lem:back.exposed} establishes
that the back faces of such bands are exposed, and hence a strip is
available. Any vertical strip extending from a top to a bottom
edge of the back face may be used.

\begin{figure}[htbp]
\centering
\includegraphics[width=\linewidth]{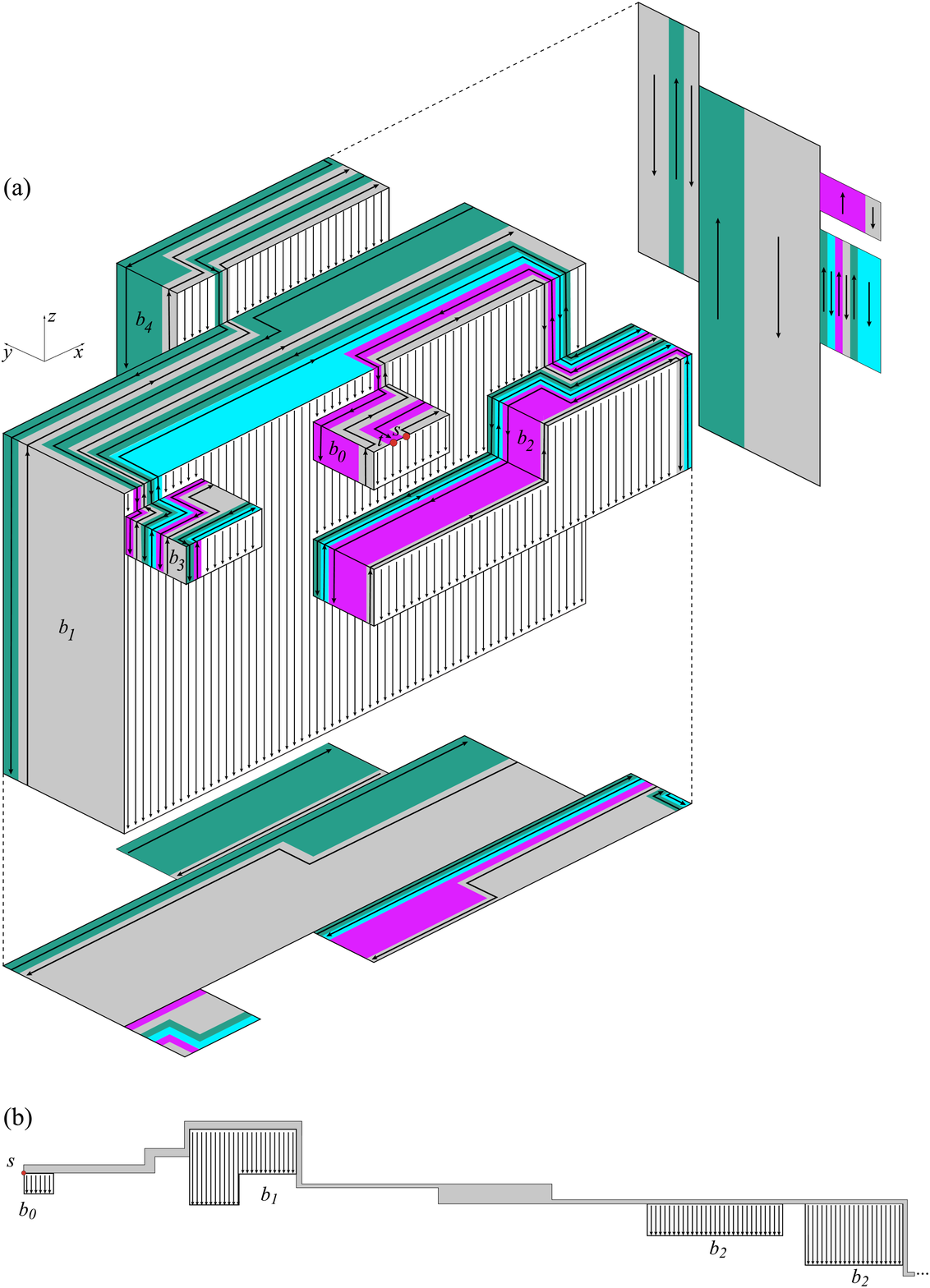}
\caption{(a) Four-block example; (b) Prefix of unfolding
(not to same scale), with front face pieces labeled.} 
\label{fig:complete.example}
\end{figure}

Figure~\ref{fig:complete.example} shows a complete unfolding example,
with the spiral path already thickened so that it covers the bands.
The unfolding begins at point $s$ on front box $b_0$.  It spirals
clockwise around $b_0$ to the $z$-beam that takes it to back child
$b_1$. Once on $b_1$, it begins following an alternating path to reach
the $z$-beams to front children $b_2$ and $b_3$. After spiraling
around $b_2$ and $b_3$ it makes one complete cycle around $b_1$ and
then follows the $z$-beam to back child $b_4$. It spirals around
$b_4$, turns around on its back face, and then tracks back through
$b_4$, around $b_3$, through front children $b_2$ and then $b_1$, and
finally around $b_0$ to point $t$.  A small portion of the
staircase-like unfolding of this example is shown in 
Figure~\ref{fig:complete.example}(b).

Front and back faces of $O$ are partitioned and attached to $\xi$
according to the illumination model described in
Section~\ref{sec:front.back}, with one modification---here both top
{\it and} bottom edges of each rim illuminate downward lightrays.  The
front and back face pieces are attached to their illuminating rim
segments.  This is illustrated by arrows on the front faces in
Figure~\ref{fig:complete.example}. Bottom edges must illuminate light
because lower bands may block rays from higher bands (which cannot
occur with extrusions).  For example, bands $b_0$, $b_2$, and $b_3$
block lightrays from $b_1$, but rays from their bottom edges
illuminate the front face pieces below them.  This method is guaranteed to
illuminate all front and back faces since a ray shot upward from any
front or back face point will hit a top or bottom edge of a rim before
leaving the surface of $O$. The rim it hits is the one that
illuminates it and the one to which its piece is attached.

With the exception of these changes, 
the algorithm remains identical to the one described
in Section~\ref{sec:unfolding.extrusions}. A summary of the algorithm,
with changes to the earlier procedure UNFOLD-EXTRUSION($O$) (from
Section~\ref{sec:unfolding.extrusions}) marked in italic, is provided
below.

\begin{center}\begin{minipage}{\linewidth}
\hrule\hfill \\
\noindent UNFOLD($O$) \hrule\hfill
\begin{enumerate}
\squeezelist
\item Partition $O$ into bands with $xz$ parallel planes $Y_0, Y_1, \ldots$
through each vertex (Section ?).
\item {\em Determine connecting $z$-beams for all pairs of $z$-visible
bands} (Section~\ref{sec:connecting.beams}.)
\item Select root band $b_0$ adjacent to $Y_0$ and {\em compute unfolding tree $T_U$
with root $b_0$} (Section~\ref{sec:unf.tree}).
\item For each band $b$ encountered in a preorder traversal of $T_U$
\begin{itemize}
\item[3.1] LABEL-FRONT-CHILDREN($b$).
\item[3.2] LABEL-BACK-CHILDREN($b$) (Section~\ref{sec:LR}).
\end{itemize}
\item Determine $\xi$ = SPIRAL-PATH($b_0$) (as in Section~\ref{sec:recursive.spiral}, but
{\em moving up and down $z$-beams}).
\item Thicken $\xi$ to cover all bands in $T_U$ (Section~\ref{sec:thicken}).
\item Hang front and back faces off $\xi$ (as in Section~\ref{sec:front.back} but illuminating light from top {\it and bottom rim edges}).
\end{enumerate}
\hrule\hfill
\end{minipage}\end{center}

\section{Worst Case}
\label{sec:worst}

\hide{
\xxx[JOR]{Rough draft by JOR on the worst case.}

\xxx[RF]{I reorganized/modified this some when adding the 
upper bound argument.}
}

The thinness of the spiral path is determined by the number of
parallel paths on any face, which we call the \emph{path density} on
that face.  If the maximum density is $k$, then the path can be at
most $1/k$-th of the face width ($y$-extent).  
We say a band $b_i$ is
\emph{visited} each time the spiral enters the band, alternates back
and forth between its children, turns around using the last back child
(or back face of $b_i$ if there are no back children), and alternates
between its children in reverse order, and finally exits $b_i$. If
there are $m$ parallel paths on a face after the first visit, then
after $v$ visits there are $vm$ parallel paths, since each subsequent
visit tracks alongside a path laid down during an earlier visit. For
example, Figures~\ref{fig:box1}--\ref{fig:box34} show a single box
visited once, and there are $4$ parallel paths on the top face: two
from $s$ to turnaround, and two back to $t$.  Figure~\ref{fig:xi.3d}
shows that with the exception of the turnaround box ($b_{10}$ in this
example) boxes at depth $d=1$ ($b_1 - b_9$) are visited twice,
doubling their path densities to $8 = 2 \times 4$.

\begin{figure}[htbp]
\centering
\includegraphics[width=0.75\linewidth]{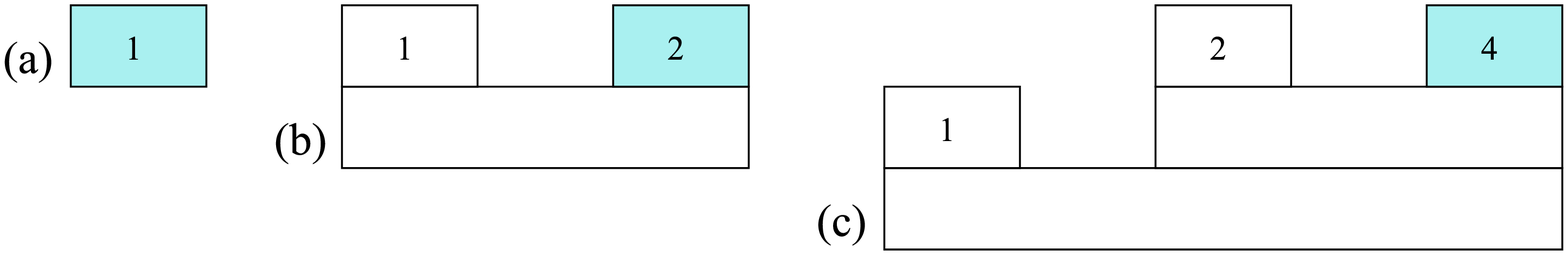}
\caption{One back child has path density $2^{d}4$.}
\label{fig:2-to-n}
\end{figure}

We now use this doubling property to construct an example that has
path density $2^{\Omega(n)}$, where $n$ is the number of vertices of
the polyhedron.  We use a skewed binary tree, as illustrated by the
sequence of extrusions (viewed from above) in
Figures~\ref{fig:2-to-n}a-c. In each case, the spiral starts on the bottom
box heading to the right. Each non-leaf box has two
back children; the right child is visited first and the
left one is the turnaround box.
The number of times each leaf box is
visited is marked.
One of the children at depth $d$ (shaded in the figure) is visited
$2^d$ times and has a path density of $2^{d}4$.  A depth-$d$ tree of
this structure contains $2d+1$ boxes total, and so can be realized by
a polyhedron with $n = 8(2d+1)$ vertices.  Thus $d = \Omega(n)$, and
the shaded child has a path density of $2^{\Omega(n)}$.  We conclude
that the spiral path may need to be as thin as $\varepsilon =
1/2^{\Omega(n)}$ times the smallest $y$-extent of any face of the
polyhedron.


We now establish a density upper bound of $2^{O(n)}$.  Observe that
each time a band is visited, its children are visited at most twice,
and therefore $2^d$ is an upper bound on the number of visits for a
band at depth $d$.  It can be that the most dense band is not a leaf,
but a band having many children. As Figure~\ref{fig:xi.exit} makes
clear, the number of parallel paths, $m$, laid out on each visit
depends on the number of children a band has, due to the alternation
back and forth to each child. Noting that both $m$ and $d$ are $O(n)$,
we can conclude that the path density is bounded by $m2^d =
O(n)2^{O(n)}$, which is still $2^{O(n)}$.


\begin{theorem}
The path density from the described algorithm is
$2^{\Theta(n)}$, and so $\varepsilon = 1/2^{\Theta(n)}$, in the worst case.
\end{theorem}

\section{Conclusion}
\label{sec:Conclusion}
We have established that every orthogonal polyhedron
of genus zero may be unfolded.
We believe that our algorithm can be extended
to handle orthogonal polyhedra with genus $\ge 1$. One idea is to
treat holes as being blocked by virtual membranes, unfold according
to our genus-$0$ algorithm, and then compensate for the virtual
faces. However, a number of details in such an algorithm would need
careful handling.


A natural extension of our algorithm would be to construct a
$k \times k$-refined grid unfolding, for constant $k$. Although our
algorithm fundamentally relies on $\varepsilon$-thin strips, a mix
of our current unfolding techniques with the ones employed
in~\cite{dfo-umt-05} to reverse the direction of the unfolding, may
help achieve this extension.

Finally, our spiraling technique so relies on the orthogonal
structure of the polyhedra that it seems difficult to use it in
resolving the open problem of whether every polyhedron may be
unfolded.

\bibliographystyle{alpha}
\bibliography{MT}

\newcommand{\etalchar}[1]{$^{#1}$}
\begin{thebibliography}{BDD{\etalchar{+}}98}

\bibitem[BDD{\etalchar{+}}98]{bddloorw-uscop-98}
T.~Biedl, E.~Demaine, M.~Demaine, A.~Lubiw, J.~O'Rourke, M.~Overmars,
  S.~Robbins, and S.~Whitesides.
\newblock Unfolding some classes of orthogonal polyhedra.
\newblock In {\em Proc. 10th Canad. Conf. Comput. Geom.}, pages 70--71, 1998.

\bibitem[DEE{\etalchar{+}}03]{deeho-vusm-03}
E.~D. Demaine, D.~Eppstein, J.~Erickson, G.~W. Hart, and J.~O'Rourke.
\newblock Vertex-unfoldings of simplicial manifolds.
\newblock In Andras Bezdek, editor, {\em Discrete Geometry}, pages 215--228.
  Marcel Dekker, 2003.
\newblock Preliminary version appeared in \emph{18th ACM Symposium on
  Computational Geometry}, Barcelona, June 2002, pp. 237-243.

\bibitem[DFO05]{dfo-umt-05}
M.~Damian, R.~Flatland, and J.~O'Rourke.
\newblock Unfolding {Manhattan} towers.
\newblock In {\em Proc. 17th Canad. Conf. Comput. Geom.}, pages 204--207, 2005.

\bibitem[DFO06]{dfo-gvuop-06}
M.~Damian, R.~Flatland, and J.~O'Rourke.
\newblock Grid vertex-unfolding orthogonal polyhedra.
\newblock In {\em Proc. 23rd Symp. on Theoretical Aspects of Comp. Sci.}, pages
  264--276, February 2006.
\newblock \emph{Lecture Notes in Comput. Sci.}, Vol. 3884, Springer.

\bibitem[DIL04]{dil-gvuo-04}
E.~D. Demaine, J.~Iacono, and S.~Langerman.
\newblock Grid vertex-unfolding of orthostacks.
\newblock In {\em Proc. Japan Conf. Discrete Comp. Geom.}, pages 76--82, 2004.
\newblock \emph{Lecture Notes in Comput. Sci.}, Vol. 3742, Springer.

\bibitem[DO04]{do-op03-04}
E.~D. Demaine and J.~O'Rourke.
\newblock Open problems from {CCCG} 2003.
\newblock In {\em Proc. 16th Canad. Conf. Comput. Geom.}, 2004.

\bibitem[DO05]{do-sfucg-05}
E.~D. Demaine and J.~O'Rourke.
\newblock A survey of folding and unfolding in computational geometry.
\newblock In J.~E. Goodman, J.~Pach, and E.~Welzl, editors, {\em Combinatorial
  and Computational Geometry}, pages 167--211. Cambridge University Press,
  2005.

\end{thebibliography}
\end{document}